\documentclass{article}

\usepackage{multirow}
\usepackage{hyperref}
\usepackage{textcomp}
\usepackage{spconf,amsmath,graphicx}
\usepackage{booktabs}
\usepackage{hyperref}

\title{ZERO-SHOT EMOTION TRANSFER FOR CROSS-LINGUAL SPEECH SYNTHESIS}
\name{ Yuke Li$^{1,}$$^\text{‡}$, Xinfa Zhu$^{1,}$$^\text{‡}$, Yi Lei$^{1}$, Hai Li$^{2}$, Junhui Liu$^{2}$, Danming Xie$^{2}$, Lei Xie$^{1,*}$\thanks{‡ Authors contributed equally to this work.}\thanks{* Corresponding author.}\vspace{-7pt}}
\address{$^1$Audio, Speech and Language Processing Group (ASLP@NPU), School of Computer Science, \\ Northwestern Polytechnical University, Xi'an, China\\
  $^2$iQIYI Inc., Chengdu, China\\}
\begin{document}
\begin{sloppypar}

%
\maketitle
\begin{abstract}


Zero-shot emotion transfer in cross-lingual speech synthesis aims to transfer emotion from an arbitrary speech reference in the source language to the synthetic speech in the target language. Building such a system faces challenges of unnatural foreign accents and difficulty in modeling the shared emotional expressions of different languages. Building on the DelightfulTTS~\cite{delightfultts} neural architecture, this paper addresses these challenges by introducing specifically-designed modules to model the \textit{language-specific} prosody features and \textit{language-shared} emotional expressions separately. Specifically, the language-specific speech prosody is learned by a non-autoregressive predictive coding (NPC) module~\cite{liu2020non} to improve the naturalness of the synthetic cross-lingual speech. The shared emotional expression between different languages is extracted from a pre-trained self-supervised model HuBERT with strong generalization capabilities. We further use hierarchical emotion modeling to capture more comprehensive emotions across different languages. Experimental results demonstrate the proposed framework's effectiveness in synthesizing bi-lingual emotional speech for the monolingual target speaker without emotional training data\footnote{Speech samples: \href{https://ykli22.github.io/ZSET/}{https://ykli22.github.io/ZSET/}}.

\end{abstract}
\section{Introduction}
\label{sec:intro}
\begin{figure}[htp]
    \centering
    \includegraphics[width=8.5cm]{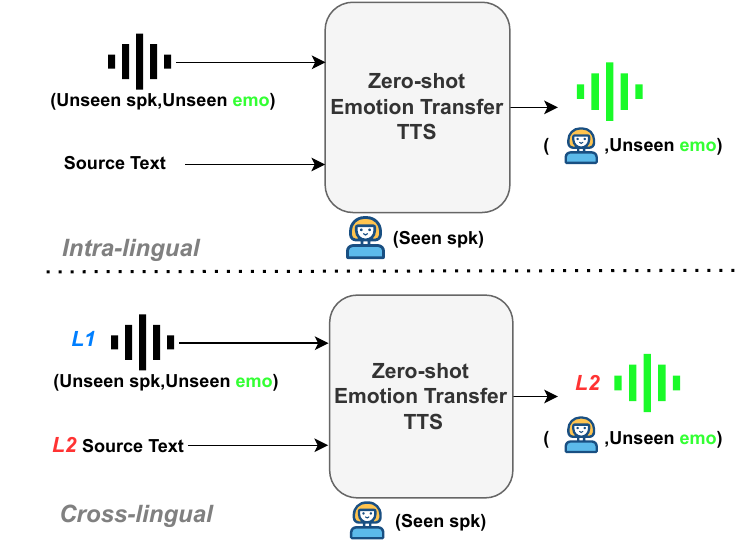}
    \caption{The overview of zero-shot emotion transfer in intra-lingual scenario (the upper part) and cross-lingual scenario (the bottom part). Notably, L1 and L2 in the figure mean any two languages.
    }
    \label{fig:galaxy}
\end{figure}



Thanks to the rapid development of deep learning, recent neural text-to-speech (TTS) technology has experienced remarkable advancements, producing natural and expressive speech~\cite{taylor2009text}~\cite{Tacotron}~\cite{Fastspeech2}. Recent studies have broadened TTS to diverse scenarios, including multi-speaker, multi-emotion, and multi-lingual. Zero-shot emotion transfer TTS has also significantly developed, which aims to transfer emotion from an arbitrary reference to synthetic speech for the target speaker without emotional expressions~\cite{chen2021adaspeech}~\cite{huang2022generspeech}. This paper focuses on a more challenging scenario -- zero-shot emotion transfer for \textit{cross-lingual} TTS.
Specifically, as shown in Figure 1, as compared to the intra-lingual scenario, in the cross-lingual scenario, the speech reference comes from another language different from the target synthetic speech. Zero-shot emotion transfer for cross-lingual TTS benefits various fields, such as movie dubbing, audiobook production, and speech translation. Nevertheless, it's quite challenging to achieve rich prosody and high naturalness in the extremely complicated cross-lingual~\cite{wu2008cross}~\cite{chen2019cross} and multi-emotion~scenarios~\cite{cooper2020zero}.

To the best of our knowledge, the research on emotion transfer across languages is still in its infancy, and zero-shot emotion transfer across languages has not been explored yet.  This is probably because of the following challenges.
First, a native L1 language speaker is exposed to a \textit{foreign accent} problem when he/she speaks the L2 language due to the different linguistic systems between L1 language and L2 language. The foreign accent can largely influence the listening perceptual experience~\cite{zhou2023accented}~\cite{flege1995second}~\cite{best1994emergence}. Second, accent and emotional expressions are both reflected in speech prosody patterns (such as pitch, energy, and duration) and heavily \textit{entangled} in speech. So how to distinguish and sufficiently disentangle the accent-related and emotion-related prosodic variations is the key factor in cross-lingual emotion transfer. Third, it's challenging to effectively extract an unseen emotional expression from the reference signal and generate natural speech in another language following the accurate emotional expression. 

In this paper, we propose a novel zero-shot cross-lingual emotional TTS approach based on the DelightfulTTS~\cite{delightfultts}. To address the issue of entangled accent and emotion in speech representations, the paper introduces a Non-Autoregressive Predictive Coding (NPC) module and a hierarchical emotion encoder to factorize prosody variations into \textit{language-specific accent features} and \textit{language-shared emotional representations}, which are recomposed to produce the natural cross-lingual emotional speech.
Based on a self-supervised approach with the masked reconstruction of speech representations, the NPC module aims to learn the prosodic patterns across different languages based on the local dependencies of speech to address the foreign accent problem. In addition, the hierarchical emotion encoder leverages HuBERT~\cite{hubert} to extract a speaker-independent representation. The HuBERT model is trained by a large amount of unlabeled data with a strong generalization ability, which can significantly improve the performance of zero-shot emotion transfer. Furthermore, considering that various acoustic and linguistic properties tend to be encoded in different layers of the self-supervised models~\cite{pasad2021layer}~\cite{baevski2020wav2vec}~\cite{chang2021exploration}, we obtain emotion representation at different granularities, aiming to capture and convey the subtle differences in emotions between different languages more comprehensively. 
In both objective and subjective evaluations, the proposed approach outperforms the baseline approach, achieving more natural and expressive emotion-transfer speech synthesis across languages. Ablation studies also demonstrate the effectiveness of the proposed method.

\section{Related Work}
\label{sec:related work}
\subsection{Cross-lingual speech synthesis}
\label{ssec:subhead}
Cross-lingual text-to-speech has drawn much attention recently. As each speaker in the training set speaks only one language, there is a heavy entanglement of linguistic content and speaker identity in speech. This induces a foreign accent problem -- The synthetic speech has a strong foreign accent. To address this problem, language ID has been employed to control the language-related rhythm ~\cite{zhang2019learning}~\cite{nekvinda2020one}~\cite{delightfultts}. Given the target language id in inference, these approaches can simply separate the linguistic information from speaker information to obtain natural cross-lingual synthetic speech. However, as different languages possess distinct phonemes and phonetic features, simply using language id is insufficient to remove the foreign accent completely in the synthetic speech. 

Domain adversarial training has been proposed to decouple speakers' timbre and language in some studies~\cite{xin2021disentangled}~\cite{nekvinda2020one}. Zhang et al.~\cite{zhang2019learning} introduce adversarial losses to separate the speaker identity from the speech content. Apart from decoupling,  Kim et al.~\cite{kim2023crossspeech}, and Staib et al.~\cite{staib2020phonological} have explored the use of a shared cross-language alphabet, reducing the reliance on language-specific graphemes. 
These methods effectively address the foreign accent problem in cross-lingual speech synthesis, but it still remains an unsolved problem in cross-lingual emotion transfer due to the heavy entanglement between language, speaker timbre, and emotion factors. 
\subsection{Emotion transfer speech synthesis}
\label{ssec:subhead}
Among the existing techniques for emotion transfer, the mainstream approach involves obtaining emotion representations through a joint-trained reference encoder~\cite{8683623}~\cite{9003859} ~\cite{zhang2023iemotts} ~\cite{huang2022generspeech}, which are then incorporated into the acoustic model for speech synthesis. The typical reference encoders include Global Style token~\cite{gst}, VAE~\cite{vae}, etc., which generate emotion representations from Mel-spectrum to participate in the acoustic modeling where labeled data are needed for supervision. To capture complex emotions more comprehensively, some researchers~\cite{lei2022msemotts}~\cite{joint}~\cite{li2021towards} have explored extracting multi-level or multi-scale emotional representations from reference audio or text.

The demand for personalized speech generation has been continuously increasing, and the application of emotion transfer in unseen scenarios has gained widespread attention.
One intuitive approach is data-driven, where acoustic models are trained with a large amount of speech data with various emotional styles~\cite{chen2021adaspeech}~\cite{paul2020speaker}. However, it incurs high costs for data collection. Some researches focus on the method to improve generalization ability through disentanglement. Huang et al.~\cite{huang2022generspeech} achieve this goal by disentangling representations into domain-specific parts and domain-invariant parts.

These emotion transfer approaches achieve admirable performance in intra-lingual speech synthesis. However, how to effectively transfer unseen emotion from the source speaker in L1 language to the target speaker in L2 language is still a brand new challenge for cross-lingual emotion transfer TTS.

\section{Proposed Method}
\begin{figure*}[!htp]
\centering
\includegraphics[width=\linewidth]{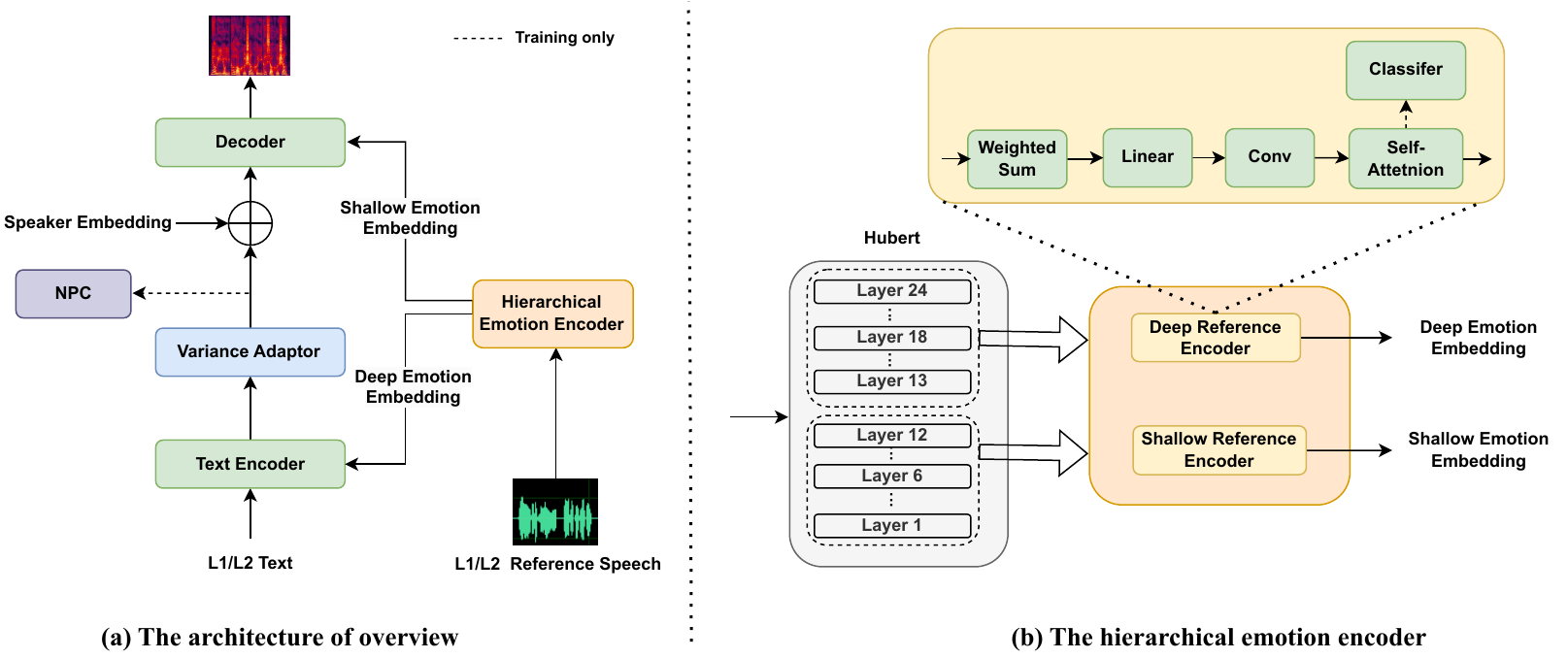}
\caption{Architecture of our system (a) and the hierarchical emotion encoder (b).}
\label{fig:y equals x}
\end{figure*}
The proposed method extends the DelightfulTTS system\cite{delightfultts} by incorporating an NPC module and a hierarchical emotion module, as shown in Figure 2. The hierarchical emotional module is responsible for learning hierarchical emotion representations using HuBERT. These representations are then passed to the encoder and decoder to synthesize emotional speech. And the NPC module is exclusively employed during the training stage, which models prosody for all languages in a self-supervised way. Finally, we use HiFiGAN~\cite{kong2020hifi} to generate speech from the Mel spectrogram.
\subsection{Backbone model}
\label{ssec:subhead}
The backbone of zero-shot cross-lingual emotion transfer speech synthesis follows the DelightfulTTS~\cite{delightfultts}, which consists of the encoder, variance adaptor, and decoder. Both the encoder and decoder are constructed using Conformer Blocks, which combine self-attention and convolutional feed forward module. This architecture is designed to capture both global-related information using self-attention and local-related information using the convolutional network. In the process, the encoder takes phoneme sequences as input and encodes them into hidden features. The variance adaptor models different variation information. The length regulator in the variance adaptor aligns the hidden text representation mixed with pitch and energy from phone level to frame level according to the duration. Finally, the decoder takes the length-regulated hidden sequence as input and generates the target Mel spectrogram. To support multi-speaker speech synthesis, speaker embedding is obtained through a look-up table, which enables the model to generate speech with different speaker characteristics.

In order to improve the performance of emotion transfer, we leverage conditional layer normalization (CLN) conditioned on hierarchical emotion representations. CLN has been proven effective in adaptive TTS~\cite{chen2021adaspeech}. By substituting all layer normalizations of the decoder with CLN conditioned on hierarchical emotion representations, we enhance the model's ability to capture and express emotional nuances during synthesis.
Overall, this framework provides the foundation for zero-shot cross-lingual emotion transfer, incorporating components to enable accurate and expressive emotion transfer across languages.

\subsection{NPC for speech representation learning}
\label{ssec:subhead}

Due to the significant differences in the phonetic features and prosodic features across different languages, the prosody information modeled in cross-lingual TTS may become inaccurate or incomplete, resulting in accent-related issues. To tackle this challenge, we incorporate the NPC module into our model, which is a self-supervised method and learns a speech representation only relying on local dependencies of speech. 

As shown in Figure 3, by using Mask Language modeling, the NPC module takes the frame sequence $ (x_{1},x_{2},x_{3}...x_{n}) $ as input after masking with $\textit{MASK}_{size}=3$, i.e. $(x_{t-1},x_{t},x_{t+1})$. This is done to prevent the model from directly copying the values of adjacent points $(x_{t-1},x_{t+1})$ as the predicted value for the current time step $x_{t}$, which avoids situations where the predicted value does not carry any information gain relative to the input $x_{t}$. The input passes through several layers of convolutional blocks, and each layer of convolutional blocks will also mask out the intermediate frames layer by layer and finally add them up as hidden features. After the hidden features pass through the VQ layer and the linear layer, the final prediction sequence is obtained. We optimize the prediction sequence and the surface feature $x_{t}$ for all time steps by L1 Loss, which is described as:
\begin{equation}
\mathcal{L}_{\mathrm{npc}} = L1 (Linear(VQ(x_{masked}),x)
\end{equation}
where the $x_{masked}$ and x separately are the ground truth frame sequence and masked frame sequence after Conv Blocks. $Linear$
and $VQ$ means the full-connect layer and the VQ layer.

Through mask language modeling, the NPC module can predict the target frame without seeing itself by relying on local context information. This idea of learning contextual embedding based on the local neighbors in the sequence has been found useful in speech representation learning. In our model, we use this network to supervise the output of the variance adaptor, which helps the model better capture the prosody patterns of all languages and address the foreign accent problem.
\begin{figure}[htp]
    \centering
    \includegraphics[width=8cm]{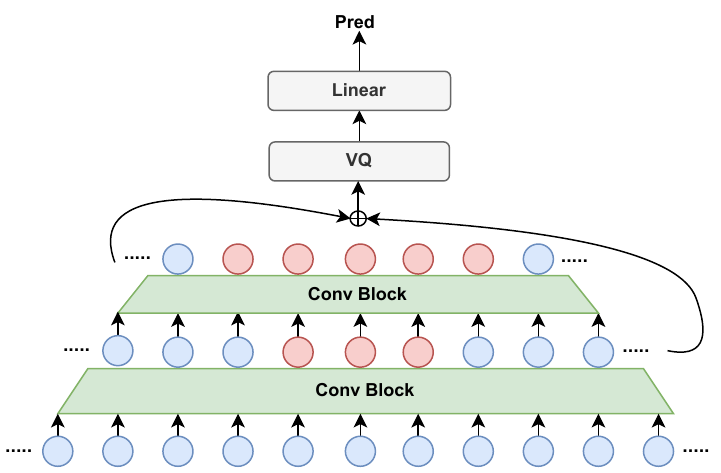}
    \caption{The architecture of the NPC module. The red circle represents the masked frame sequence.}
    \label{fig:galaxy}
\end{figure}
\subsection{Hierarchiacl emotional modeling}
\label{ssec:subhead}
Zero-shot emotion transfer involves a wide array of complex and diverse emotions. In order to enhance the model's generalization ability on different languages, we employ HuBERT~\cite{hubert}, which is pre-trained on 10,000 hours of Chinese speech data from the WenetSpeech dataset in an unsupervised manner~\footnote{Hubert:\href{https://github.com/TencentGameMate/chinese_speech_pretrain/}{https://github.com/TencentGameMate/chinese\_speech\_pretrain/}}. 
Some studies~\cite{pepino2021emotion}~\cite{macary2021use} have already discovered the self-supervised pre-trained model can extract emotional information to achieve emotion recognition. Since different layers of HuBERT contain distinct information, we adopt a strategy of utilizing HuBERT representations from different layers to extract hierarchical emotional features. This approach aims to improve the performance of emotion transfer compared to relying solely on a single layer. The shallower layers of HuBERT often capture more fine-grained information, while the higher layers provide stronger semantic information but have a limited perception of details. As a result, we extracted shallow emotion embedding using HuBERT representations from layers 1 to 12 and deep emotion embedding using representations from layers 13 to 24.

As shown in Figure 2, the emotion encoder includes a shallow reference encoder and a deep reference encoder, which have the same structure. The reference encoder weights and sums the representations of HuBERT's different layers to extract emotion representation. The structure of reference encoders follows the Mel-style-encoder in Meta-StyleSpeech~\cite{min2021meta}. To obtain the emotion embeddings, which are 24-layer embeddings of HuBERT, a piece of 200-frame audio samples is randomly intercepted as input of HuBERT. The emotion embeddings are then fed into a fully-connected layer, followed by a convolution layer with residual connections to capture the sequence information.
Multi-head attention is used to encode global information. Finally, shallow emotion embedding is added to the decoder input, and deep emotion embedding is integrated into the text encoder. During training, we pre-train this module using part of the emotion labeling data to guide the unlabeled data for emotion modeling. Deep and shallow emotion representations conduct semi-supervised learning through cross-entropy loss.

\begin{equation}
\mathcal{L}_{\mathrm{emo}} = CE(EmoID_{pred},EmoID_{truth})
\end{equation}
where $EmoID_{pred}$ is predicted emotion id got from this module and $EmoID_{truth}$ is groundtruth emotion id.

\subsection{Training and inference}
\label{ssec:subhead}
During training, we first pre-train the hierarchical emotion encoder using L1 language data, which utilizes a small amount of labeled emotional data for supervised guidance. Subsequently, we conduct joint training of the acoustic model and emotional encoder using the bilingual dataset. This joint training process fine-tunes the emotion encoder, enabling the establishment of a language-shared emotional space. The acoustic model is optimized by minimizing the L1 distance of the predicted Mel-spectrum, duration, pitch, and energy. The training objective is
\begin{equation}
\begin{aligned}
\mathcal{L}_{\mathrm{total}} =& \mathcal{L}_{\mathrm{mel}}+\mathcal{L}_{\mathrm{pitch}}+\mathcal{L}_{\mathrm{energy}}
+\mathcal{L}_{\mathrm{dur}} \\ 
&+\mathcal{L}_{\mathrm{npc}}+\mathcal{L}_{\mathrm{emo}}
\end{aligned}
\end{equation}
where $\mathcal{L}_{\mathrm{mel}}$ is the MSE loss to measure the similarity between predicted and ground-truth Mel spectrogram. $\mathcal{L}_{\mathrm{pitch}}$, $\mathcal{L}_{\mathrm{energy}}$, and $\mathcal{L}_{\mathrm{dur}}$ are the L1 losses between predicted pitch/energy/duration and the ground-truth pitch/energy/duration respectively. 
For inference, the model can use any unseen emotional speech in any language as the reference speech for the zero-shot, and the text from another language is used as the input. Certainly, using paired speech-text data can lead to better results in the emotional transfer process. The hierarchical emotion encoder extracts language-shared emotion embeddings. At the same time, the translated text is processed to be the phoneme sequence and input into the text encoder. By leveraging the emotion embeddings and aligning them with the translated text, the model ensures that the generated target language audio maintains the desired emotional characteristics present in the reference audio, thus achieving an effective emotion transfer.

\section{Experiments}
\label{sec:typestyle}

\subsection{Datasets}
\label{ssec:subhead}
\begin{table*}[!htbp]
\caption{Comparison of the proposed model with M3, SCLN-GST, and ablation models regarding emotion similarity and speaker naturalness MOS with confidence intervals of 95\% under 2 voice conversion scenarios. NMOS denotes naturalness MOS, and EMOS denotes emotion similarity MOS. SPKCOS means speaker cosine similarity. A higher value means better performance.}
\vspace{6pt}
\label{tab:mos}
\setlength{\tabcolsep}{3mm}
\centering
\resizebox{\linewidth}{!}{
 \renewcommand{\arraystretch}{1.2}
\begin{tabular}{lccccccccc}
\hline
\multirow{2}{*}{model} & \multicolumn{4}{c}{CN$\xrightarrow{}$TH}  &  & \multicolumn{4}{c}{TH$\xrightarrow{}$CN}  \\ \cline{2-5} \cline{7-10}
                       & EMOS $\uparrow$ & NMOS $\uparrow$ & SPKCOS $\uparrow$ & CER $\downarrow$ &  & EMOS $\uparrow$ & NMOS $\uparrow$ & SPKCOS $\uparrow$ & CER $\downarrow$ \\ \hline
GT                    & -    & -    & -      & 0.059   &  & -    & -    & -  & 0.002   \\
M3                    & 2.17$\pm$0.14 & 2.04$\pm$0.12 & 0.768 & 0.360   &  & 2.24$\pm$0.12 & 2.03$\pm$0.15 & 0.542 & 0.146    \\ 
SCLN-GST              & 3.51$\pm$0.10 & 3.73$\pm$0.09 & 0.831  & 0.090   &  & 3.15$\pm$0.11 & 3.44$\pm$0.13& 0.784 & 0.022   \\ \hline
\textbf{Proposed}    & \textbf{3.95$\pm$0.09} & \textbf{4.03$\pm$0.08} & \textbf{0.879}  & 0.071   &  &\textbf{3.94$\pm$0.10}  &\textbf{3.96$\pm$0.08} &\textbf{0.838} & 0.017  \\ 
\hspace{1em}-NPC      & 3.77$\pm$0.08 & 3.54$\pm$0.13   & 0.855  & 0.076   &  &3.51$\pm$0.13   &3.45$\pm$0.14 & 0.794 & 0.022   \\ 
\hspace{1em}-Deep emo  & 3.42$\pm$0.11 & 3.98$\pm$0.10 & 0.816   & 0.070   &  & 3.23$\pm$0.13   & 3.66$\pm$0.10 & 0.810 & 0.017   \\
\hspace{1em}-Shallow emo  & 3.36$\pm$0.10 & 4.01$\pm$0.12 & 0.834   & \textbf{0.068}   &  & 3.20$\pm$0.11   & 3.78$\pm$0.09 & 0.821 & \textbf{0.016}   \\ \hline
\end{tabular}
}
\end{table*}
We use two internal speech corpus as the training set. Chinese corpus contains 23-hour speeches from 7 Chinese speakers. Four speakers have 6 emotions, including happiness, anger, sadness, surprise, disgust, hate, and fear, while other speakers do not exhibit any specific emotions in their speech. Thai corpus contains 20-hour Thai speech from 8 Thai speakers, where only one speaker has apparent emotional expression. 

All speech is down-sampled to 24kHz, and 80-dimensional Mel-spectrogram extracted with 12.5 ms frameshift and 50 ms frame length are used. For front-end processing, we use the front-end trained by the HMM model to process Chinese text to phoneme sequence, tone, prosody, and segment information, and Thai text is processed by the open-source front-end ``thainlp"~\cite{pythainlp} to perform word segmentation and convert word to phoneme, and get its tone at the same time.
       
\subsection{Model configuration}
\label{ssec:subhead}
We adopt SCLN-GST~\cite{gst-tts} and M3~\cite{shang2021incorporating} as the compared systems for zero-shot cross-lingual emotion transfer speech synthesis.
\begin{itemize}
\item \textbf{SCLN-GST}: A cross-speaker emotion transfer framework~\cite{gst-tts} with Global Style Token as reference encoder, while the model learns speaker embeddings by SCLN blocks.
\item \textbf{M3}: A multi-speaker multi-style multi-language speech synthesis system~\cite{shang2021incorporating}, which incorporates a speaker conditional variational encoder for adversarial training to decouple the speaker information. Additionally, style information is extracted by a fine-grained style encoder.
\item \textbf{Proposed}: The proposed cross-lingual cross-speaker zero-shot emotion transfer framework.
\end{itemize}               
In our implementation of the proposed approach, the backbone of the model follows the settings of DelighfulTTS, and CLN consists of 2 fully connected layers. The NPC structure consists of 4 masked conv blocks with the masked size is 5. The reference encoder in hierarchical emotion encoder follows the structure of the Mel-style encoder proposed by Meta-StyleSpeech~\cite{min2021meta}.

For the training setup, the Adam optimizer is used to optimize the network with an initial learning rate of 0.002 and a batch size of 32. It is trained for 400K steps. We use the same dataset to train the HifiGAN ~\cite{kong2020hifi} vocoder with a total number of steps of 500k.
\subsection{Subjective evaluation}
\label{ssec:subhead}

The performance of three models on zero-shot emotion transfer is evaluated in terms of emotion similarity and naturalness with the Mean Opinion Score (MOS) test. The test set consists of 30 unseen Chinese emotional speech samples and 30 unseen Thai emotional speech samples. Taking the Chinese to Thai (CN to TH) scenario as an example, we prepare the reference Chinese speech and corresponding translated Thai texts to generate the Thai samples for two neutral Chinese speakers. We invite ten native Chinese speakers and five native Thai speakers to participate in the subjective evaluation. The listeners are asked to evaluate the emotional similarity between the synthetic and source speech on a scale from 1 to 5. 

As shown in Table 1, the proposed method in this paper achieves the best performance in emotion similarity and naturalness. The high emotion similarity demonstrates the effectiveness of hierarchical emotion modeling. Besides, the proposed method achieves the best speech naturalness, indicating its ability to generate cross-lingual speech and mitigate foreign accents. In addition, SCLN-GST achieves lower emotion similarity and naturalness. This could be attributed to the fact that SCLN-GST is a single-language emotion transfer model, and its performance in cross-lingual emotion transfer is generally limited. On the other hand, M3, which is a cross-lingual style transfer model, exhibited inferior performance when it comes to emotion transfer.

\subsection{Objective evaluation}
\label{ssec:subhead}
In objective evaluations, we use speaker cosine similarity and character error rate (CER) of the synthesized speech to measure the speaker similarity and overall speech quality. We use the pre-trained ECAPA-TDNN ~\cite{desplanques2020ecapa} to extract the speaker embedding and compute the cosine similarity between the source speech and the synthetic speech. In addition, we use MMS ~\cite{pratap2023mms} to evaluate the CER of synthetic Thai speech and Whisper~\cite{radford2022whisper} finetuned by Chinese data to evaluate the CER of synthetic Chinese speech, respectively.

As shown in Table 1, the test results show that the proposed model obtains the highest speaker cosine similarity and achieves a CER close to the ground truth audio, indicating its clarity and intelligibility. However, SCLN-GST and M3 exhibit higher CER in speech recognition than the proposed model, suggesting that both models may face challenges in accurately preserving linguistic clarity while transferring emotions in Chinese and Thai languages.

\subsection{Ablation study}
\label{ssec:subhead} 
The original version of delightful TTS cannot perform emotion transfer, so to demonstrate the validity of our proposed approach, we conduct an ablation study to examine the impact of two key components: the NPC module and hierarchical emotion modeling. Specifically, we compare our full model's performance with three variants: first without the NPC module(-NPC), the second using only shallow-level emotion representations instead of the hierarchical ones(-Deep emo), and the third using only deep-level emotion representations instead of the hierarchical ones(-Shallow emo).

As shown in Table 1, removing the NPC structure leads to a small decrease in naturalness, indicating that the NPC module plays a critical role in addressing the foreign accent problem and improving the naturalness of synthetic speech. On the other hand, when relying on single-level emotion representations, we notice a significant decline in expressiveness. 
Moreover, we find that the model without shallow emotion representations exhibits lower emotion similarity than the model without deep emotion representations. However, it demonstrates lower CER and higher speaker similarity. 
These results suggest that shallow emotion representations contribute to emotional expressiveness while deep emotion representations contain more semantic information, benefiting natural speech and accurate pronunciations.
By employing hierarchical emotion modeling, we accurately capture subtle emotional differences at different levels, thereby improving the emotional similarity, naturalness, and expressiveness of the generated speech.

\section{Conclusion}
\label{sec:illust}

This paper proposes a cross-lingual hierarchical emotion modeling framework to achieve zero-shot emotion transfer in cross-lingual TTS. The proposed model utilizes an NPC module for self-supervised speech representation learning to solve foreign accent problems. Furthermore, we design the hierarchical emotion encoder to learn a language-shared emotional representation. The experimental result shows that the proposed approach can effectively transfer emotion from the source to the target language, synthesizing the foreign accent-free and expressively emotional speech.


\end{sloppypar}
\end{document}